\documentclass[a4paper]{jpconf}
\usepackage{graphicx}
\begin{document}
\title{An Unfolded Quantization for Twisted Hopf Algebras}

\author{Francesco Toppan}

\address{
Dep. TEO, CBPF, Rua Dr. Xavier Sigaud 150, cep 22290-180, Rio de Janeiro (RJ), Brazil.
}

\ead{toppan@cbpf.br}

\begin{abstract}
In this talk I discuss a recently developed ``Unfolded Quantization Framework". It allows to introduce a Hamiltonian Second Quantization based on a Hopf algebra endowed with a coproduct satisfying, for the Hamiltonian, the physical requirement of being a primitive element. The scheme can be applied to theories deformed via a Drinfel'd twist.  I discuss in particular two cases: the abelian twist deformation of a rotationally invariant nonrelativistic Quantum Mechanics (the twist induces a standard noncommutativity)
and the Jordanian twist of the harmonic oscillator. In the latter case the twist induces a Snyder
non-commutativity for the space-coordinates, with a pseudo-Hermitian deformed Hamiltonian.\par
The ``Unfolded Quantization Framework" unambiguously fixes the non-additive effective interactions in the multi-particle sector of the deformed quantum theory. The statistics of the particles is preserved even in the presence of a deformation.
\end{abstract}

\section{Introduction}

In this talk I will summarize the main results, recently appeared in four separate papers in J. Math. Phys.
\cite{{cct},{ckt}, {cckt}, {ckt2}}, concerning a new Hamiltonian quantization framework for
(twisted) Hopf algebras, which is applicable to noncommutative nonrelativistic Quantum Mechanics. 

This paper is intended to provide a self-contained quick introduction to the addressed problem and its proposed solutions. It will report the main relevant results obtained so far.\par
 Let us start with the formulation of the problem.  Hopf algebras are a natural framework to
deal with deformed theories. Indeed, a Drinfel'd twist preserves the Hopf algebra axioms while deforming its structures/costructures. Among the Hopf algebra (co)structures, the undeformed coproduct admits a natural physical interpretation. The undeformed coproduct of a primitive element $\Omega$  is expressed through 
\begin{eqnarray}\label{prim}
\Delta(\Omega)& =&\Omega\otimes{\bf 1}+{\bf 1}\otimes \Omega.
\end{eqnarray} For our purposes we can consider here the Hopf algebra defined
on the Universal Enveloping Algebra ${\cal U}({\cal G})$ of a Lie algebra ${\cal G}$. A primitive element is a Lie algebra element (therefore $\Omega\in {\cal G}$).\par
It is quite natural to apply this undeformed scheme to a system of free, non-interacting quantum
systems: a chain of non-interacting harmonic oscillators, a chain of non-interacting spins, and so forth.  The free Hamiltonian $H$ should therefore be regarded as a primitive element, with the undeformed coproduct translating, into the Hopf algebra setting, the additivity of energy for the
non-interacting multi-particle state (the total energy being recovered as the sum of the single-particle energy levels).\par
These considerations are quite straightforward. It is also straightforward to realize that a Drinfel'd twist can induce a deformation of the free theory. The real issue comes when we try to be specific
and implement this framework in a concrete setting. It has been observed several times in the literature (see the discussion in \cite{cct}) that the standard quantization framework based on the Heisenberg
algebra ${\cal H}$  (with creation and annihilation operators acting on a Fock space) fails to provide a Hamiltonian which satisfies (\ref{prim}). Let, for a harmonic oscillator,
set $H=\omega a^\dagger a$. Then, the creation and annihilation operators $a^\dagger, a$ satisfy
$\Delta (a^\dagger) =a^\dagger\otimes {\bf 1}+{\bf 1}\otimes a^\dagger$ and
$\Delta (a) =a\otimes {\bf 1}+{\bf 1}\otimes a$. On the other hand
$\Delta (H) \neq H\otimes {\bf 1}+{\bf 1}\otimes H$. The reason for that is that $H$ is an element of the Heisenberg enveloping algebra and not one of the Heisenberg algebra generators.\par
This simple example shows us that we cannot use, as we would have naively expected, ${\cal U}({\cal H})$, the Universal Enveloping Algebra of the Heisenberg algebra, as the undeformed Hopf algebra setting for the Second Quantization of the harmonic oscillator.\par
Mathematically, there is nothing wrong of course with the Hopf algebra defined on ${\cal U}({\cal H})$. It is a mathematically consistent Hopf algebra. What ${\cal U}({\cal H})$ fails to provide is a consistent, {\em
physical} interpretation for the undeformed coproduct (in accordance with (\ref{prim})). If we 
insist in using it and apply to the Second Quantization of the harmonic oscillator, we simply get
wrong results.\par
It is clear, from the previous considerations, which strategy could be found to overcome this problem. Since formula (\ref{prim}) applies to primitive elements, then we need to realize the undeformed free Hamiltonian as a primitive element. One possibility \cite{cct} is offered by the so-called
Wigner's quantization \cite{wigner} (mostly known in the literature, see \cite{palev}, as Wigner's oscillators).
In the Wigner's approach the Hamiltonian of the harmonic oscillator, in its symmetrized, Weyl form, can be expressed as an anticommutator. We have $H=\{F_+,F_-\}$, where $F_\pm$ are two odd generators which, in Wigner's case, play the role of the creation/annihilation operators of the standard approach. $F_\pm$ belong to the odd sector ${\cal  G}_{1}$ of the $osp(1|2)$ superalgebra whose even sector ${\cal G}_0$ is isomorphic to $sl(2)$ ($H, E_\pm\in {\cal G}_0$).
$H$ corresponds to the Cartan's element of $osp(1|2)$. \par
The Fock's space is replaced, in the Wigner's setting, by a lowest weight representation of
$osp(1|2)$. The lowest weight vector $|\lambda>$  ($F_-|\lambda>=0$, $H|\lambda>=\lambda |\lambda>$) corresponds to the Fock's vacuum, while the lowest weight $\lambda$ can be interpreted as a vacuum energy.  If $|\lambda>$ is assumed to be bosonic, the lowest weight representation (spanned by the vectors ${F_+}^n|\lambda>$) contains bosonic vectors for even values of $n$ and fermionic vectors for odd values of $n$. The standard Hilbert space of the harmonic oscillator is recovered \cite{cct} by setting a specific value for the vacuum energy 
$\lambda$ and by projecting out the fermionic sector.\par
In the Wigner's approach, $H$ is a Cartan element of $osp(1|2)$ and therefore a primitive element
of the superalgebra. The Wigner's approach to the Second Quantization of the harmonic oscillator \cite{cct},  based on the Hopf algebra structure defined for the ${\bf Z}_2$-graded Universal Enveloping Algebra ${\cal U}(osp(1|2))$ is consistent, both mathematically and physically.\par
In dealing with more complicated Hamiltonians (not necessarily coinciding with harmonic oscillators and their extensions) we would like to have at disposal a more flexible picture
for a Second Quantization based on Hopf algebra.
This alternative, more flexible picture, introduced at first in \cite{ckt} and further elaborated in \cite{{cckt},{ckt2}}, goes under the name of ``Unfolded Quantization". It will be discussed in the following.

\section{The Unfolded Quantization}

The Unfolded Quantization requires the preliminary identification of an abstract, dynamical Lie algebra ${\cal G}_d$ which contains the Hamiltonian among its generators. The identification of
${\cal G}_d$, that is which operators should belong to ${\cal G}_d$ and regarded as primitive
elements, is imposed by physical considerations and the constraints put on the problem that we are investigating. Let us consider as an example a rotationally invariant three-dimensional nonrelativistic system.  The angular momentum ${\vec L}$ is a primitive element,
while ${\vec L}^2$ is not. This has to do with the fact that the coproduct
\begin{eqnarray}
\Delta ({\vec L}^2)&=& \Delta({\vec L})\cdot\Delta({\vec L}) = {\vec L}^2\otimes {\bf 1} + {\bf 1}\otimes {\vec L}^2 + 2{\vec L}\otimes {\vec L}
\end{eqnarray}
encodes the notion that ${\vec L}^2$ is not an additive operator. Indeed, for a composite system we have the identity
$({\vec L}_{1+2})^2=({\vec L}_1+{\vec L}_2)^2$.\par
To be specific let us discuss the example of the three-dimensional harmonic oscillator.
Its Heisenberg algebra is given by the generators 
$\hbar, x_i, p_i$ ($i=1,2,3$). Its non-vanishing commutation relations are
\begin{eqnarray}\label{heiscomm}
\relax [x_i,p_j]&=& i\delta_{ij}\hbar.
\end{eqnarray}
 We can define as its dynamical Lie algebra ${\cal G}_d$, the one given by the generators
$\hbar, x_i, p_i, L_i, H,K,D$, together with its set of abstractly defined non-vanishing commutation relations
\begin{eqnarray}\label{ocomm}
\relax [x_i,p_j]&=&i\delta_{ij}\hbar,\nonumber\\
\relax [x_i, L_j] &=&i\epsilon_{ijk}x_k,\nonumber\\
\relax [p_i, L_j]&=& i\epsilon_{ijk}p_k,\nonumber\\
\relax [x_i, H] &=& 2ip_i,\nonumber\\
\relax [x_i,D]&=& ix_i,\nonumber\\
\relax [p_i, D] &=& -i p_i,\nonumber\\
\relax [p_i,K]&=& - 2 i x_i,\nonumber\\
\relax [H, D]&=& -2i H,\nonumber\\
\relax [H,K]&=& -4 i D,\nonumber\\
\relax [D,K]&=& - 2i K,
\end{eqnarray}
determined by the Heisenberg commutators (\ref{heiscomm}) via the identifications
\begin{eqnarray}
L_i &=& \frac{1}{\hbar}\epsilon_{ijk}x_jp_k,\nonumber\\
H&=& \frac{1}{\hbar} {\vec p}^2,\nonumber\\
D&=& \frac{1}{2\hbar}({\vec x}{\vec p}+{\vec p}{\vec x}),\nonumber\\
K&=& \frac{1}{\hbar}{\vec x}^2.
\end{eqnarray}
Within ${\cal G}_d$, the Hamiltonian ${\bf H}$ of the harmonic oscillator is expressed as $
{\bf H}=H+K$.\par
As mentioned in the Introduction, the Hopf algebra structure relevant for physical purposes is defined for ${\cal U}({\cal G}_d)$.\par
For more general Hamiltonians whose potential is no longer quadratic, an infinite number
of generators need to be introduced to consistently define the dynamical Lie algebra ${\cal G}_d$. For such cases (unlike the harmonic oscillator case) ${\cal G}_d$ becomes an infinite-dimensional Lie algebra.\par
The Hopf algebra structure of ${\cal U}({\cal G}_d)$ neatly encodes the Second Quantization.
On the other hand, we can perform the Second Quantization without even referring to its Hopf algebra structure just like, in a different context, we can perform sums of trigonometric functions
without referring to the Hopf algebra structure of real functions closed under addition
(the formula $sin(x+y)=sin(x) cos(y)+cos( x) sin( y)$ being traslated, in the Hopf algebra setting,
as $\Delta (sin) = sin\otimes cos + cos\otimes sin$).\par
The real advantage of the introduction of a Hopf algebra structure lies in the fact that it
allows controlling the deformations. In the following we apply the Unfolded Quantization to some
selected examples of Drinfel'd twist \cite{{drin85},{drin88}}.

\section{Deformations: the abelian twist}

The ${\cal U}({\cal G}_d)$ algebra can be deformed via a twist ${\cal F}\in {\cal U}({\cal G}_d)\otimes{\cal U}({\cal G}_d)$ satisfying the cocycle condition. The abelian twist is defined to be
\begin{eqnarray}\label{abtwist}
\cal{F} &=& \exp\left(i \alpha_{ij} p_i\otimes p_j\right), \quad \alpha_{ij}=-\alpha_{ji}.
\end{eqnarray}
It is abelian since  $[p_i,p_j]=0$ and well-defined due to the fact that the $p_i$ momenta are among the generators of ${\cal G}_d$.
\par
The twist induces a deformation ($g\mapsto g^{\cal F}$) for the ${\cal G}_d$ generators, with $g^{\cal F}$ belonging to ${\cal U}({\cal G}_d)$. 
The generators which commute with $p_i$ remain undeformed. Among the deformed generators we have
\begin{eqnarray}
 x_i^{\mathcal{F}} &=& x_i - \alpha_{ij} p_j \hbar, \nonumber \\
 K^{\mathcal{F}} &=& K - \alpha_{ij}x_ip_j + \frac{\alpha_{jk}\alpha_{jl}}{2!}p_kp_l \hbar.
\end{eqnarray} 
The deformation of the position operators $x_i$ corresponds to the Bopp shift.\par
In the so-called ``hybrid formalism" (see \cite{cct}) one can link the abelian twist to the (constant) noncommutativity, through the relations
\begin{eqnarray}
[x_i^{\cal F}, x_j^{\cal F}]&=& i\Theta_{ij},
\end{eqnarray}
where the constant operator $\Theta_{ij}$ is given by
\begin{eqnarray}
\Theta_{ij}&=&2\alpha_{ij}\hbar^2.
\end{eqnarray}
For single-particle operators the knowledge of the deformed generators, together with their commutators and their action on a module $V$ which possesses the structure of a Hilbert space,
is sufficient to quantize the system. For multi-particle operators the extra-structure of the (deformed) coproduct plays a role. The deformed $2$-particle operator associated with the
deformed generator $g^{\cal F}$ is constructed by applying $\Delta^{\cal F}(g^{\cal F})\in 
{\cal U}^{\cal F}({\cal G}_d)\otimes  {\cal U}^{\cal F}({\cal G}_d)$ to the Hilbert space $V\otimes V$. The twist ${\cal F}$ (\ref{abtwist}), applied to $V\otimes V$, corresponds to the unitary operator $F$.
Since 
\begin{eqnarray}
\Delta^{\cal F}(g^{\cal F}) &=& {\cal F}\cdot \Delta (g^{\cal F})\cdot  {\cal F}^{-1},\nonumber\\
\end{eqnarray}
with $\Delta (g^{\cal F})$ the undeformed coproduct, we end up that the operators
$\widehat{\Delta^{\cal F}}(g^{\cal F})$, $\widehat{\Delta}(g^{\cal F})$, acting on $V\otimes V$, are unitarily equivalent:
\begin{eqnarray}
\widehat{\Delta^{\cal F}}(g^{\cal F}) &=& F\cdot \widehat{\Delta}(g^{\cal F})\cdot F^{-1}.
\end{eqnarray}
This feature also applies for $n$-particle operators with $n\geq 3$. \par
It is convenient to introduce the symbol `` ${\widehat{}}$ " when we need to make the distinction
between an element $\Omega$ of the (tensor product of the) Universal Enveloping Lie Algebra and
its action ${\widehat \Omega}$ on a module. Therefore, ${\bf H}^{\cal F}\in {\cal U}^{\cal F}({\cal G}_d)$ while ${\widehat {\bf H}^{\cal F}}:V\rightarrow V$.\par
Let us express the deformation parameter $\alpha_{ij}$ through
\begin{eqnarray}
\alpha_{ij}&=&\epsilon_{ijk}\frac{\alpha_k}{Z}.
\end{eqnarray}
Without loss of generality we can set ${\vec\alpha}=(0,0,\alpha_3=\alpha)$. 
Once applied the abelian twist-deformation (\ref{abtwist}) the deformed Hamiltonian
reads as
\begin{equation}
\mathbf{H}^\mathcal{F} = H+K-\alpha (xp_y- yp_x)+\frac{\alpha^2}{2}\hbar(p_x^2+p_y^2).
\end{equation}
The undeformed coproduct of the deformed Hamiltonian reads
\begin{eqnarray}\label{defcop3}
\Delta(\mathbf{H}^\mathcal{F})&=&\mathbf{H}^\mathcal{F}\otimes\mathbf{1}+\mathbf{1}\otimes\mathbf{H}^\mathcal{F}+\alpha(y\otimes p_x+p_x\otimes y
-x\otimes p_y-p_y\otimes x) \nonumber \\
&+&\frac{\alpha^2}{2}\sum_{i=1}^2(2p_i\hbar\otimes p_i+2p_i\otimes p_i\hbar+p_i^2\otimes\hbar+\hbar\otimes p_i^2).
\end{eqnarray}
It is symmetric under particle exchange. Therefore, the particles behave as ordinary bosons even in the presence of the deformation.
One should also note that the deformed two-particle Hamiltonian is no longer additive due to the extra terms which depend on $\alpha$.  Even if no longer additive, the coassociativity of the coproduct guarantees in any case the associativity of the deformed Hamiltonian. Indeed, for three-particle states, we have the equality  
\begin{equation}
(id\otimes\Delta)\Delta(\mathbf{H}^\mathcal{F})=(\Delta\otimes id)\Delta(\mathbf{H}^\mathcal{F})\equiv\Delta_{(2)}(\mathbf{H}^\mathcal{F}),
\end{equation}
where, explicitly 
\begin{eqnarray}\label{delta2}
\Delta_{(2)}(\mathbf{H}^\mathcal{F})&=&\mathbf{H}^\mathcal{F}\otimes
\mathbf{1}\otimes\mathbf{1}+\mathbf{1}\otimes\mathbf{H}^\mathcal{F}
\otimes\mathbf{1}+\mathbf{1}\otimes\mathbf{1}\otimes\mathbf{H}^\mathcal{F} \nonumber\\
&&+\alpha(\mathbf{1}\otimes y\otimes p_x+y\otimes\mathbf{1}\otimes p_x+y\otimes p_x\otimes\mathbf{1})\nonumber\\
&&+\alpha(\mathbf{1}\otimes p_x\otimes y+p_x\otimes\mathbf{1}\otimes y+p_x\otimes y\otimes\mathbf{1})\nonumber\\
&&-\alpha(\mathbf{1}\otimes x\otimes p_y+x\otimes\mathbf{1}\otimes p_y+x\otimes p_y\otimes\mathbf{1})\nonumber\\
&&-\alpha(\mathbf{1}\otimes p_y\otimes x+p_y\otimes\mathbf{1}\otimes x+p_y\otimes x\otimes\mathbf{1})\nonumber\\
&&+\alpha^2\sum_{i=1}^2[\mathbf{1}\otimes p_i\hbar\otimes p_i+p_i\hbar\otimes p_i\otimes\mathbf{1}+p_i\hbar\otimes p_i\otimes\mathbf{1}\nonumber\\
&&+\mathbf{1}\otimes p_i\otimes p_i\hbar+p_i\otimes p_i\hbar\otimes\mathbf{1}+p_i\otimes p_i\hbar\otimes\mathbf{1}\nonumber\\
&&+\hbar\otimes p_i\otimes p_i+p_i\otimes p_i\otimes\hbar+p_i\otimes p_i\otimes\hbar\nonumber\\
&&+\frac{1}{2}(\mathbf{1}\otimes \hbar\otimes p_i^2+\hbar\otimes p_i^2\otimes\mathbf{1}+\hbar\otimes p_i^2\otimes\mathbf{1}\nonumber\\
&&+\mathbf{1}\otimes p_i^2\otimes \hbar+p_i^2\otimes \hbar\otimes\mathbf{1}+p_i^2\otimes \hbar\otimes\mathbf{1})].
\end{eqnarray}
The deformed two-particle energy $E_{12}^{\cal F}$ can be expressed as
\begin{eqnarray}\label{defaddit}
E_{12}^{\cal F} &=& E_1^{\cal F}+E_2^{\cal F}+\Omega_{12},
\end{eqnarray}
where $E_i^{\cal F}$ ($i=1,2$) are the single-particle energies and $\Omega_{12}$ is an effective interaction term. Therefore we have at least two possible interpretations for the above results. Either we regard $\Omega_{12}$ as an interaction or we regard (\ref{defaddit}) as describing a system of free (albeit deformed) particles, with $\Omega_{12}\neq 0$ as a measure of deformation.\par
The associativity is expressed by the three-particle formula
\begin{eqnarray}\label{intter2}
E_{123}^{\cal F} &\equiv & E_{(12)3}^{\cal F}=E_{1(23)}^{\cal F}=E_{1}^{\cal F}+E_{2}^{\cal F}+E_{3}^{\cal F}+\Omega_{12}+\Omega_{23}+\Omega_{31} +\Omega_{123},
\end{eqnarray}
with $\Omega_{123}$ recovered from the $\Omega_{ij}$'s.\par
It should be stressed the crucial role of the coproduct in unambiguously determine the ``interacting term" $\Omega_{12}$.\par
The formulas (\ref{defcop3}) and (\ref{delta2}) are equalities in the tensor products of the Universal Enveloping Lie algebras ${\cal U}^{\cal F}({\cal G}_2)\otimes {\cal U}^{\cal F}({\cal G}_2)$
and ${\cal U}^{\cal F}({\cal G}_2)\otimes {\cal U}^{\cal F}({\cal G}_2)\otimes {\cal U}^{\cal F}({\cal G}_2)$, respectively.
To get the $2$-particle, $3$-particle (and in general multi-particle) Hamiltonian we have to apply on the $V\otimes \ldots\otimes V$ 
multi-particle Hilbert space (with $\hbar$ mapped into the identity operator). 
It should be stressed that the computation of the interacting term ${\widehat \Omega}_{12}$, made possible by the application of the Unfolded Quantization framework,  goes beyond the results of
\cite{{kijanka},{scholtz}}.

\section{Deformations: the Jordanian twist and the Snyder noncommutativity}

The formalism of the Unfolded Quantization can be applied to another example of Drinfel'd twist,
the Jordanian deformation of $sl(2)$ \cite{dv, ohn, og}. It is defined by the twist
\begin{equation}
{\cal F}=\exp\left(-iD\otimes\sigma\right),
\end{equation}
where  $\sigma=\ln(\mathbf{1}+\xi H)$. The parameter $\xi$ is dimensional and is taken as a real, positive number. It can be applied to twist the Universal Enveloping Algebra $\mathcal{U}({\cal G}_d)$, with ${\cal G}_d$ defined in (\ref{ocomm}). As before the twist induces a deformation $g\mapsto g^{\cal F}$ on the generators of ${\cal G}_d$. Explicitly, the deformed generators are
\begin{eqnarray}
x_i^{\cal F}&=&x_i e^{\frac{\sigma}{2}}\nonumber\\
p_i^{\cal F}&=&p_i e^{-\frac{\sigma}{2}}\nonumber\\
H^{\cal F}&=&H e^{-\sigma}\nonumber\\
K^{\cal F}&=&K e^{\sigma},
\end{eqnarray}
the others remaining undeformed. \par
The deformed Hamiltonian of the harmonic oscillators reads as
\begin{equation}
\mathbf{H}^{\cal F}=H^{\cal F}+K^{\cal F}=H e^{-\sigma}+K e^{\sigma}.
\end{equation}
In the hybrid formalism \cite{ckt} the commutator of the deformed position variables yields the Snyder noncommutativity \cite{snyder}:
\begin{eqnarray}
[x_i^{\cal F}, x_j^{\cal F}]&=&-\frac{i\xi}{2}(x_i^{\cal F} p_j^{\cal F}-x_j^{\cal F} p_i^{\cal F}).
\end{eqnarray}

The other nonvanishing commutators are
\begin{eqnarray}
\relax [x_i^{\cal F}, p_j^{\cal F}]&=&i\hbar\delta_{ij}+\frac{i\xi}{2}p_i^{\cal F} p_j^{\cal F} \nonumber\\
\relax [x_i^{\cal F}, D^{\cal F}]&=&\frac{i}{2}(x_i^{\cal F} -\xi x_i^{\cal F} H^{\cal F} ) \nonumber\\
\relax[x_i^{\cal F}, H^{\cal F}]&=&ip_i^{\cal F}(\mathbf{1}-\xi H^{\cal F}) \nonumber\\
\relax[x_i^{\cal F}, K^{\cal F}]&=&-\frac{\xi}{2}x_i^{\cal F}\left(\mathbf{1}+\frac{\xi}{2} H^{\cal F}\right)+i\xi(K^{\cal F} p_i^{\cal F}+D^{\cal F} x_i^{\cal F})  \nonumber\\
\relax[p_i^{\cal F}, D^{\cal F}]&=&-ip_i^{\cal F}\left(\mathbf{1}-\frac{\xi}{2} H^{\cal F}\right) \nonumber\\
\relax[p_i^{\cal F}, K^{\cal F}]&=&-i(x_i^{\cal F}+\xi p_i^{\cal F} D^{\cal F})+\frac{\xi^2}{4}p_i^{\cal F} H^{\cal F} \nonumber\\
\relax[D^{\cal F}, H^{\cal F}]&=&iH^{\cal F}(\mathbf{1}-\xi H^{\cal F}) \nonumber\\
\relax[D^{\cal F}, K^{\cal F}]&=&-iK^{\cal F}(\mathbf{1}-\xi H^{\cal F}) \nonumber\\
\relax[K^{\cal F},H^{\cal F}]&=&2iD^{\cal F}(\mathbf{1}+\xi H^{\cal F})+2\xi H^{\cal F}-2\xi^2 (H^{\cal F})^2.
\end{eqnarray}
As in the abelian twist case, deformed and undeformed coproducts are unitarily equivalent
for all $n$-particle states. The particles remain bosonic even in the presence of the deformation.
The associativity of the multi-particle Hamiltonian is guaranteed by the coassociativity of the coproduct.

The main difference with respect to the abelian case is that now the deformed Hamiltonian is no
longer Hermitian. On the other hand, it belongs to the so-called $\eta$-type class of pseudo-Hermitian Hamiltonians, satisfying 
\begin{equation}
{\bf H}^{{\cal F}\dagger}=\eta{\bf H}^{\cal F}\eta^{-1},
\end{equation}
with $\eta$ a Hermitian operator. For the Jordanian twist we have
\begin{equation}
\eta = e^\sigma=\mathbf{1}+\xi H.
\end{equation}
A consistent quantum mechanics can be made with such Hamiltonians, see \cite{kato, most}.
For that it is sufficient to define an inner product $<<,>>$ under which $\mathbf{H}^{\cal F}$ is self-adjoint. Obviously this deformed inner product needs to satisfy some conditions, for instance, positiveness.  

This different inner product is related to the usual one by
\begin{equation}\label{inner product}
<< \psi,\phi >>=< \psi, \eta \phi >.
\end{equation}
Note that $\eta$ is a Hermitian, linear and invertible operator, and is analogous to a metric in a usual finite-dimensional vector space. 

Although as a vector space the Hilbert space endowed with the $\eta$-deformed inner product is isomorphic to the original one, as Hilbert spaces this is not true. Thus, we denote the Hilbert space with the $\eta$-deformed inner product as $\tilde {\cal H}$. If $\eta$ is positive definite, the new inner product will also be so. 
 
Under this inner product we have  
\begin{eqnarray}
\nonumber << \psi, \mathbf{H}^{\cal F} \phi >> &=& < \psi ,e^{\sigma} \mathbf{H}^{\cal F} \phi > \\ 
\nonumber &=& < e^{\sigma} \mathbf{H}^{\cal F} e^{-\sigma} e^{\sigma} \psi, \phi >\\
\nonumber &=&< \mathbf{H}^{\cal F}\psi, e^{\sigma}\phi >\\
&=& << \mathbf{H}^{\cal F} \psi, \phi >>.
\end{eqnarray}
Therefore, under the deformed inner product the Hamiltonian becomes self-adjoint. Operators such as ${\bf H}^{\cal F}$, which are self-adjoint under the $\eta$-deformed inner product, are called \emph{$\eta$-pseudo-Hermitian} operators \cite{most}. 

It remains to be seen if the spectrum of the $\eta$-pseudo-Hermitian Hamiltonian is real. This will be true because the formal square root of $\eta$, $\rho$ such that $\rho^2=\eta$, is a unitary transformation $\rho : \tilde {\cal H} \rightarrow {\cal H}$, even though it is a Hermitian operator in ${\cal H}$ . 

To see this, consider a linear transformation $T: W \rightarrow V$ between two finite-di\-men\-sional vector spaces $W$ and $V$. Its adjoint will then be a transformation $T^{\ddagger}: V^{\ast} \rightarrow W^{\ast}$, where $V^{\ast}$ and $W^{\ast}$ are duals to the original vector spaces. Then we have, by definition,
\begin{equation}
< w,T^{-1}v >_W=<T^{-1 \ddagger}w,v >_V,
\end{equation}
where $<,>_W$ is the inner product in $W$ and $<,>_V$ is the one in $V$. 

In our notation for the inner products of ${\cal H}$ and $\tilde{\cal H}$ we have, for $\rho=\exp{{\frac12 \sigma}}$,
\begin{equation}
<< \tilde \psi ,\rho^{-1} \phi >> =< \rho^{-1 \ddagger}\tilde \psi, \phi >,
\end{equation}
with $\tilde \psi \in \tilde {\cal H}$ and $\phi \in{\cal H}$. Using expression (\ref{inner product}) we  also find that
\begin{equation}
<< \tilde \psi ,\rho^{-1} \phi >> =< \rho \tilde \psi, \phi >.
\end{equation}
Therefore $\rho^{-1 \ddagger} = \rho$, so that $\rho$ is unitary when regarded as a transformation $\rho : \tilde {\cal H} \rightarrow {\cal H}$. Note that $\tilde \psi$ can also be regarded as a vector in ${\cal H}$, since ${\cal H}$ and $\tilde{{\cal H}}$ are identified as vector spaces.

A useful way of dealing with this scenario is to map all observables on $\tilde {\cal H}$ back onto ${\cal H}$ where the inner product is the usual one. This is done by 
\begin{equation}
\mathbf{H}^{\cal F} \mapsto \mathbf{H}^{\cal F}_{\rho} = \rho \mathbf{H}^{\cal F} \rho^{-1}.
\end{equation}
The new Hamiltonian will be given by 
\begin{equation}
\mathbf{H}^{\cal F}_{\rho} =  \left( 1 - \frac{\xi^2}{4} \right)H^{\cal F}+ K^{\cal F} +i \xi D,
\end{equation}
which is explicitly Hermitian since $K^{{\cal F} \dagger} = K^{{\cal F}} + 2i\xi D$. For our present consideration, $\xi$ is a small parameter. This shows that our pseudo-Hermitian Hamiltonian is related to a manifestly Hermitian Hamiltonian by a unitary transformation, and is therefore guaranteed to have a real spectrum. The transformation $\rho$ is called a pseudo-canonical transformation; the systems described by ${\bf H}^{\cal F}$ in $\tilde{\cal H}$ and ${\bf H}^{\cal F}_\rho$ in ${\cal H}$ are physically equivalent \cite{most}.

\section{Conclusions}
\par
~\par
In this talk I presented the Unfolded Quantization framework (developed in \cite{{ckt}, {cckt}, {ckt2}}) which allows performing a Second Quantization within a Hopf algebra scheme which
satisfies the physical requirements for the (undeformed) coproduct. Its basic tenet is the correct
determination of the dynamical Lie algebra ${\cal G}_d$. The Hopf algebra is defined on its
Universal Enveloping Algebra. In application to Drinfel'd-twist deformations of Hopf algebras
(such as the abelian twist that was discussed in Section {\bf 3} and the Jordanian twist that was
discussed in Section {\bf 4}), the Unfolded Quantization framework leads to non-additive effective interactions in the multi-particle sectors. These interactions, induced by the twist,
satisfy consistency condition such as the associativity (induced by the coassociativity of the coproduct) and symmetry under particle exchange. The Unfolded Quantization framework unambiguously fixes the effective interactions, allowing to go beyond the results previously obtained in the literature concerning single-particle operators (see, e.g., \cite{{kijanka},{scholtz}}
in the case of the abelian twist). \par
Applied to the Jordanian twist, the Unfolded Quantization leads to some surprises: the twist
induces a Snyder-type noncommutativity for the space coordinates and, moreover, the deformed
Hamiltonian, despite being no longer Hermitian, belongs to a well-known class of pseudo-Hermitian Hamiltonians admitting a consistent quantization scheme (see \cite{{kato},{most}}.\par
An important issue which was not mentioned in this talk concerns the role of the Noether charges in the presence of a twist-deformation. In \cite{ckt} this problem was addressed for the rotational symmetry in the presence of an abelian-twist deformation. The results are quite illuminating. The introduction of twist-deformed brackets (see \cite{asc}) allows to close the
$so(3)$ rotational algebra even in the presence of the deformation. This result, however, is only
formal since it can be proven that, apart from the free-particle Hamiltonian, no other Hamiltonian with a non-constant potential is invariant under this $so(3)$ algebra. Stated otherwise, in the presence of the (abelian) twist, the $so(3)$
generators do not belong any more to a dynamical symmetry algebra of the twist-deformed
Hamiltonian. \par
These investigations concerned so far the Second Quantization of non-relativistic quantum mechanical systes. Works devoted to the application of the Unfolded Quantization framework
to DSR (deformed special relativities) theories are in progress. Another line under current investigation concerns the application of this framework to theories at finite temperature,
defined via the Takahashi-Umezawa thermofield dynamics approach \cite{thermo}.

{\bf{Acknowledgments}}

This work received support from CNPq. It is my pleasure to acknowledge my collaborators
P. G. Castro, B. Chakraborty, R. Kullock and Z. Kuznetsova.

\end{document}